\documentclass[pra,aps,groupedaddress,twocolumn,floatfix,nofootinbib]{revtex4}
\usepackage{graphicx}
\usepackage{soul}

\newcommand{\beq}{\begin{equation}}
\newcommand{\eeq}{\end{equation}}
\newcommand{\beqa}{\begin{eqnarray}}
\newcommand{\eeqa}{\end{eqnarray}}

\newcommand{\ket}[1]{\mbox{$ | #1 \rangle $}}
\newcommand{\bra}[1]{\mbox{$ \langle #1 | $}}

\def\half{\frac{1}{2}}
\def\opone{\leavevmode\hbox{\small1\normalsize\kern-.33em1}}

\begin{document}

\title{Real Numbers are the Hidden Variables of Classical Mechanics}
\author{Nicolas Gisin \\
\it \small   Group of Applied Physics, University of Geneva, 1211 Geneva 4,    Switzerland}

\date{\small \today}
\begin{abstract}
Do scientific theories limit human knowledge? In other words, are there physical variables hidden by essence forever? We argue for negative answers and illustrate our point on chaotic classical dynamical systems. We emphasize parallels with quantum theory and conclude that the common real numbers are, de facto, the hidden variables of classical physics. Consequently, real numbers should not be considered as ``physically real" and classical mechanics, like quantum physics, is indeterministic.
\end{abstract}
\maketitle

\section{Introduction}\label{intro}
``Quantum Limits of Knowledge" is the nice title of a workshop organised in the historical seminar room at the Niels Bohr Institute in Copenhagen in May 2019. At first sight, no doubts, quantum theory imposes limits to what can be known. There are Heisenberg's uncertainty relation and - Copenhagen obliged - Bohr's complementary principle. But is it scientific to believe that scientific theories limit human knowledge? In particular, does quantum theory limit our knowledge or does it faithfully describe an indeterminated world, a world in which objects do not have determined positions, momenta and further properties? In short, should one speak of the uncertainty relation or of the indeterminacy relation?

For a realist, like myself, scientific theories describe what there is, not the limits of our knowledge. One can't simultaneously know with arbitrary precision the position and momentum of particles not because of some fancy limitations to our knowledge, but merely because particles don't have simultaneous precise positions and momenta. Nevertheless, looking for additional variables is highly interesting, because it may allow one to discover new physics. This implies that the hypothetical new variables should not be hidden, at least not hidden by essence for ever: they may be hidden today, but the interest is to find and reveal them\footnote{Accordingly, I was never interested in local hidden variables - despite this common historical terminology, but by local variables and the lack thereof.}. At least, this is the rough story. In quantum theory, things are more complex, because of the locality issue, on one side, and for historical reasons on the other side. 

What about classical mechanics? Here things seem clear. To my knowledge there has never been any suggestion to organise a workshop on ``Classical limits of knowledge". But, why not?

Consider a chaotic classical dynamical system. Such systems exhibit what is known as deterministic chaos, i.e. the dynamical equations are deterministic (as is the Schr\"odinger equation) and the entire trajectory, $\vec x(t)$, $\vec p(t)$, is fully determined by the initial conditions $\vec x(0)$, $\vec p(0)$. For chaotic systems, the leading digits of $\vec x(t)$, $\vec p(t)$ depend on far down the series digits of  $\vec x(0)$, $\vec p(0)$. Such far down the series digits are clearly inaccessible - hidden, but according to textbooks on classical mechanics $\vec x(0)$ and $\vec p(0)$ are real numbers (or vectors) and real numbers have unlimited numbers of digits, containing typically infinite information (except for computable numbers, a subset of the reals of measure zero). Hence, although our technology is limited, and will always be limited, the theory says that the entire trajectory $\vec x(t)$, $\vec p(t)$ is fully determined by the initial conditions. So, if one studies textbooks, then classical mechanics is fully deterministic. But what about the physics of classical systems?

In the next section \ref{ACM} I briefly summarize an alternative theory of classical mechanics in which initial conditions, like all parameters, are given by finite-information numbers. This alternative theory makes precisely the same predictions as standard classical mechanics, hence, the alternative theory has the same huge explanatory power as standard classical mechanics. However, the alternative theory is not deterministic when applied to chaotic dynamical systems. In section \ref{supplVarACM} we look for additional variables - hidden to the alternative theory, but whose existence can be postulated in such a way as to render the supplemented alternative theory deterministic. Actually, the hidden variables are the de-facto-inaccessible real numbers and the supplemented theory is merely standard classical mechanics! Thus, I go on in section \ref{pushingBack}, the apparent no knowledge limits of classical theory is an illusion. The illusion is due to the common usage of real numbers that classical physicists assume to be physically real, i.e. to faithfully represent something physically real, although there is and can't be any scientific reasons to trust that real numbers are faithful representations of physical entities: this habit is based on a myth. A myth some may consider convenient because it renders classical mechanics deterministic. But one may also argue that this myth unnecessarily enlarges the difference between classical and quantum theories. For sure, this myth amounts to push back all the indeterminacy present in the alternative classical theory to the inaccessible initial conditions: God played all dice at the big-bang and all the future is encoded - hidden - in the real numbers assumed to faithfully describe the initial conditions, as I discuss in section \ref{pushingBack}

\section{Alternative classical mechanics}\label{ACM}
In this section I briefly recall a pretty straightforward alternative theory to standard classical mechanics, first presented in \cite{GisinReal}. This alternative theory keeps the standard dynamical equations of classical mechanics unchanged, but all parameters, in particular the initial conditions, but also masses, charges, etc., and the time parameter are restricted to finite information numbers. In \cite{GisinReal}, the motivation is that a finite volume of space can't contain infinite information, hence, e.g., the center of mass of an object can't be faithfully represented by a typical real numbers, because typical real numbers contain infinite information. Such parameters should be described by finite information numbers. 

What precisely finite information numbers are (besides containing finite information) is not too important to the present paper. Reference \cite{FlavioNG19} elaborates on this. For sure, finite information numbers contain all computable numbers, as the information content of a computable number is the information that defines the algorithm to compute it and such information is finite. Masses, charges and similar parameters are given by computable numbers. But finite information numbers go beyond computable numbers and this provides useful representations of initial conditions in our alternative classical theory. 

A full description of finite information numbers requires intuitionistic mathematics, see, e.g., \cite{IndeterminateNumbersPosy,GisinIntuitionism}. But here it suffices to think of them as follows. Let $x$ be a finite information number. For simplicity and without loss of generality, we assume $x$ is larger than 0 and smaller than 1 and use base 2 to expand the bits of $x$:
\beq
x=b_1b_2b_3...b_n...
\eeq
In contrast to real numbers, here the $b_n$'s are not truly bits, their values are not restricted to bit values 0 and 1, but can assume any rational value between 0 and 1. Thus, strictly speaking, $x$ is not a number in the usual sense, it is a process that develops in time; in \cite{FlavioNG19} we named such $x$'s Finite Information Quantities (FIQs). Let us emphasize that such quantities are not static, their bits are not all given at once, but evolve as time passes (as ``numbers" in intuitionistic mathematics \cite{IndeterminateNumbersPosy,Kreisel,Troelstra,GisinIntuitionism}). Each $b_n$ is interpreted as the tendency, or propensity, that it eventually settles at the bit value 1. Hence $b_n=1$ means that this bit is fixed to the determined value 1 and similarly $b_n=0$ means that its value is fixed to 0. Think of the first $b_n$ taking values 0 or 1, they are determined. However, the far down the series $b_n$ are totally random, i.e. $b_n=\half$ for $n$ large enough. In-between the $b_n$'s may take any (rational) value such that the total information content of $x$ is finite:
\beqa
I(x)&=&\sum_{n=1}^\infty I(b_n) \\
&=&\sum_{n=1}^\infty (1-h(b_n)) \\
&=&\sum_{n=1}^\infty (1+b_n\log_2(b_n)+(1-b_n)\log_2(1-b_n) \nonumber\\
&<& \infty
\eeqa
Physicists are not used to think of numbers as developing in time, and indeed, our FIQs $x$ are, as said, not numbers according to classical mathematics. However, let us repeat that FIQs are perfectly valid numbers in intuitionistic mathematics, where numbers are processes that develop in time \cite{IndeterminateNumbersPosy, Kreisel, Troelstra, GisinIntuitionism}. But for this paper, physicists' intuition suffices. Indeed, physicists are used to the idea that the leading digits of physical parameters are relevant, fixed and determined, while far down the series digits are not yet relevant, possibly not yet determined\footnote{For example, as early as 1955, Born wrote: ``Statements like `a quantity x has a completely definite value' (expressed by a real number and represented by a point in the mathematical continuum) seem to me to have no physical meaning" \cite{born}.}. Whether this indeterminacy is purely epistemic or fundamental (i.e. ontological) is usually (and often wisely) left open. But here, for the purpose of this paper on limits of knowledge, we assume this indeterminacy to be fundamental. 

Let's come back to classical chaos. At small times, the leading digits of the initial conditions suffice to determine the evolution, at least for all practical purposes. However, quickly, the evolution will depend on far down the series digits. Hence, as time passes, new digits must acquire determined and fixed values. Which values is not important, since anyway the evolution is chaotic: the newly determined digits can be random (possibly with some correlations), as random as the far down the series digits of typical real numbers \cite{Borel,Chaitin,DowekRealNb13}.

Accordingly, we face a choice. Either the present state of a chaotic system reveals retroactively information about long passed initial conditions represented by real numbers, or the present results from an indeterminated state of affairs represented by FIQs. The first view is that of standard classical mechanics, the second of alternative classical mechanics. Both predict indistinguishable random evolutions of chaotic dynamical systems\footnote{Possibly sharing the same global structure, like, e.g., a strange attractor.}, and both make the same predictions for integrable dynamical systems, as in such systems only the leading digits are relevant.

\section{Supplementary variables to alternative classical mechanics}\label{supplVarACM}
In alternative classical mechanics initial conditions do not fully determine the evolution of chaotic dynamical systems. The evolution of such systems are described by quantities that are processes that develop in time. The development is usually not deterministic. Hence, in the alternative theory, the evolution of chaotic systems is truly indeterministic, in strong contrast to standard classical mechanics where the chaos is deterministic. Let us concentrate on the indeterministic alternative theory and look for supplementary variables. 

In full generality, one can always supplement indeterministic theories with additional variables such that the supplemented theory is deterministic \cite{GudderHV}. In a nutshell, it suffices to add as supplementary variables all results of all possible future measurements. Ideally, these are coded in such a way that they are not immediately accessible, though this aspect of supplementary variables is not always well defined. Before looking at alternative classical mechanics, let us consider two other examples based on quantum mechanics.

First, consider standard quantum mechanics, but restricted, for simplicity, to measurements with binary outcomes. A simple (and admittedly artificial) way to supplement this theory is to add to the standard state vector, $\psi\in\mathcal{H}$, a random number $r$ uniformly distributed in the unit interval $[0..1]$. Whenever a binary measurement is performed, write $r$ in binary form and divide it in 2 new random numbers $r_1$ and $r_2$. The bits of $r_1$ is the series of bits of $r$ at odd positions, while the bits of $r_2$ are $r$'s bits at even positions. Hence, both $r_1$ and $r_2$ are uniformly distributed in the unit interval. The rule is that the outcome of the binary measurement $P$ is $+1$ iff $r_1\leq\bra{\Psi}P\ket{\Psi}$ and the supplementary variable $r$, after the measurement, is updated to $r_2$. No doubt that this supplemented model is ad-hoc, but the point is simply to illustrated how easily one can supplement any indeterministic theory.

Second, consider Bohmian mechanics as quantum mechanics supplemented by (local) Bohmian positions and (nonlocal) dynamical equations for these Bohmian positions \cite{Bohm52}. Bohmian mechanics is clearly much more elegant, thus attractive to physicists, than our first example. But one may question whether it is fundamentally different is nature? In both cases the supplementary variables are inaccessible beyond the deterministic measurement outcomes they trigger. Anyway, here we like to concentrate on alternative classical mechanics.

Thirdly, consider alternative classical mechanics. Here, every physicists will immediately guess elegant supplementary variables: just supplement the finite information quantities by the usual real numbers. It this way, all future results of chaotic dynamical systems are encoded in presently inaccessible digits. This is such an elegant solution that almost all physicists accept it without even thinking about it. Note that most of these physicists nevertheless reject Bohmian position, despite the similarity of the two supplemented theories.

\section{Pushing back all indeterminacy to the initial conditions}\label{pushingBack}
In the previous section, we have seen that, generally, there is no way to distinguish experimentally an indeterministic theory from any of its supplemented deterministic theories. And vice-versa, all deterministic theories have an empirically equivalent indeterministic theory \cite{Werndl}. Somehow, instead of God playing dice when an undetermined outcome has to happen, God played all dies at the origin of times, e.g. at the big-bang. Note that the far down the series digits of typical real numbers are truly random, as random as quantum measurement outcomes performed on half a singlet, as has been emphasized, among others, by Borel \cite{Borel} and Chaitin \cite{Chaitin}. Accordingly, standard classical physics pushes all randomness to the initial condition and then claims the theory is deterministic. But one may argue that this is just a trick, an elegant trick not in contradiction with any empirical evidence, but a trick unsupported by any empirical evidence.

Consequently, we face a choice: either the fact that at present certain things happen and others do not is interpreted as revealing, retroactively, information about long past initial conditions, or else, we understand the present as the result of indeterminate reality, and the future as open. If we care about how we experience reality, the later option is obviously superior \cite{Dolev18,GisinTimePasses,GisinReal}.

\section{conclusion}\label{concl}
We argued that classical physics theory can quite naturally and intuitively be considered as an indeterministic theory supplemented by additional variables and that these additional variables are nothing but our familiar real numbers. This is interesting from several points of view.

First, it allows one to discuss additional variables outside the framework of quantum mechanics. Next, it shows that the infamous quantum measurement problem is not restricted to quantum theory, but is actually present in all indeterministic theories \cite{FlavioNG19}. Thirdly, it illustrates possible classical limits of knowledge.

Finally, our discussion illustrates the important role played by classical mathematics. Indeed, in classical mathematics, formalized following Hilbert's huge influence at the beginning of last century, the digits of all real numbers are assumed to be all given at once. This translates in classical physics, in particular in chaotic classical dynamical systems, by the assumption that all the future is given at once, i.e. encoded in the real-valued initial conditions. However, classical (Platonistic) mathematics is not the only form of mathematics. There is also intuitionistic mathematics, a form of mathematics much less known than classical mathematics, but a quite well developed mathematics in which numbers, in particular real numbers, are not all given at once, but are processes that develop in time. Brouwer, the father of intuitionism, named these processes choice sequences \cite{Brouwer1948,IndeterminateNumbersPosy,Kreisel,Troelstra}. Clearly, if physics is expressed in the language of intuitionistic mathematics, then one concludes very naturally that classical physics is indeterministic. The fact is that, historically, Hilbert won his debate with Brouwer (and Einstein won his debate with Bergson), hence physicists use classical/Platonistic mathematics and time was expulsed from physics. But there is no logical nor empirical  necessity for this state of affair. It illustrates what everyone speaking more than one language knows, namely that different languages make certain thoughts easier to express in one language than in another, like determinism is easier to express in the classical mathematics language and indeterminism easier in intuitionistic mathematics \cite{GisinIntuitionism}.

Finally, let us come back to the question raised in the introduction: is it scientific to believe that scientific theories limit human knowledge? Almost all colleagues would answer in the negative. However, most may conclude that scientific theories should be deterministic, as, if not, the future would be intrinsically unknowable. Here we argued on the contrary for indeterminism, as, if not, past real-valued initial conditions would be intrinsically unknowable. For me, there is nothing unscientific in claiming that the future is open, hence that there are limits to our knowledge of the future, but actual physical parameters of our scientific theories should be, at least in principle, knowable.

\small
\section*{Acknowledgment} This work profited from stimulating discussions with Flavio Del Santo and Christian W\"uthrich.\\ \\

\end{document}